%Paper: gr-qc/9412064
%From: damour@ihes.fr (Thibault Damour)
%Date: Wed, 21 Dec 94 11:03:02 +0100

\magnification=1200

{\baselineskip=14pt

\magnification=1200
\overfullrule=0pt

\vsize=175mm
\hsize=130mm

\voffset=2mm
\voffset=11mm

\def\Lc{{\cal L}}

\font\tenbb=msym10
\font\sevenbb=msym7
\font\fivebb=msym5
\newfam\bbfam
\textfont\bbfam=\tenbb \scriptfont\bbfam=\sevenbb
\scriptscriptfont\bbfam=\fivebb
\def\bb{\fam\bbfam}
\def\Rb{{\bb R}}

\def\un{{\rm 1\mkern-4mu I}}

\def\up#1{\raise 1ex\hbox{\sevenrm#1}}

\def\build#1_#2^#3{\mathrel{
\mathop{\kern 0pt#1}\limits_{#2}^{#3}}}

\centerline{\bf WAS EINSTEIN 100\% RIGHT ?\footnote{${}^*$}{\sevenrm Invited
talk given at the Seventh Marcel Grossmann Meeting on General Relativity
(Stanford, July 1994), to appear in the proceedings (World Scientific,
Singapore, 1995).}}

\vglue 1cm

\centerline{Thibault DAMOUR}

\centerline{\it Institut des Hautes Etudes Scientifiques}

\centerline{\it 91440 Bures-sur-Yvette, France}

\vglue 4mm

\centerline{and}

\vglue 4mm

\centerline{\it DARC, CNRS - Observatoire de Paris}

\centerline{\it 92195 Meudon, France}

\vglue 1cm

\centerline{ABSTRACT}

\vglue 4mm

{\leftskip=10mm
\rightskip=10mm
{\baselineskip=12pt
{\sevenrm The confrontation between General Relativity and experimental
results,
notably binary pulsar data, is summarized and its significance discussed. The
agreement between experiment and theory is numerically very impressive.
However, some recent theoretical findings (existence of non-perturbative
strong-field effects, natural cosmological attraction toward zero scalar
couplings) suggest that the present agreement between Einstein's \break theory
and experiment might be a red herring and provide new motivations for improving
the experimental tests of gravity.} \par}
\par}

\vglue 6mm

\noindent {\bf 1. Introduction}

\vglue 4mm

General Relativity can be thought of as defined by two postulates. One
postulate states that the action functional describing the propagation and
self-interaction of the gravitational field is
$$
S_{\rm gravitation} \ [g_{\mu \nu}]= {c^4 \over 16\pi \ G} \int
{d^4 x \over c} \ \sqrt{g} \ R(g). \eqno (1)
$$
A second postulate states that the action functional describing the coupling of
all the (fermionic and bosonic) fields describing matter and its electro-weak
and
strong interactions is a (minimal) deformation of the special relativistic
action functional used by particle physicists (the so called ``Standard
Model''), obtained by replacing everywhere the flat Minkowski metric $f_{\mu
\nu} = {\rm diag} (-1,+1,+1,+1)$ by $g_{\mu \nu} (x^{\lambda})$ and the partial
derivatives $\partial_{\mu} \equiv \partial / \partial x^{\mu}$ by
$g$-covariant derivatives $\nabla_{\mu}$. [With the usual subtlety that one
must also introduce a field of orthonormal frames, a ``vierbein'', for writing
down the fermionic terms]. Schematically, one has
$$
S_{\rm matter} \ [\psi ,A,H,g] = \int { d^4 x \over c} \ \sqrt{g} \ \Lc_{\rm
matter}, \eqno (2{\rm a})
$$
$$
\Lc_{\rm matter} = -{1 \over 4} \sum {1 \over g_*^2} \ {\rm tr} (F_{\mu
\nu} \ F^{\mu \nu} ) - \sum \overline{\mathstrut \psi} \ \gamma^{\mu} \ D_{\mu}
\
\psi
$$
$$
\qquad \quad-{1 \over 2} \ \vert D_{\mu} \ H \vert^2 - V(H) -\sum y \
\overline{\mathstrut \psi} \ H \ \psi , \eqno (2{\rm b})
$$
where $F_{\mu \nu}$ denotes the curvature of a $U(1)$, $SU(2)$ or $SU(3)$
Yang-Mills connection $A_{\mu}$, $F^{\mu \nu} =g^{\mu \alpha} \ g^{\nu \beta} \
F_{\alpha \beta}$, $g_*$ being a (bare) gauge coupling constant; $D_{\mu}
\equiv
\nabla_{\mu} +A_{\mu}$; $\psi$ denotes a fermion field (lepton or quark, coming
in various flavours and three generations); $\gamma^{\mu}$ denotes four Dirac
matrices such that $\gamma^{\mu} \ \gamma^{\nu} + \gamma^{\nu} \ \gamma^{\mu} =
2 g^{\mu \nu} \ \un_4$, and $H$ denotes the Higgs doublet of scalar fields,
with $y$ some (bare Yukawa) coupling constants.

\medskip

Einstein's theory of gravitation is then defined by extremizing the total
action functional,
$$
S_{\rm tot} \ [g,\psi ,A,H] = S_{\rm gravitation} \ [g] + S_{\rm matter} \
[\psi
,A,H,g] . \eqno (3)
$$

Although, seen from a wider perspective, the two postulates (1) and (2) follow
from the unique requirement that the gravitational interaction be mediated only
by massless spin-2 excitations [1], the decomposition in two postulates is
convenient for discussing the theoretical significance of various tests of
General Relativity. Let us discuss in turn the experimental tests of the
coupling of matter to gravity (postulate (2)), and the experimental tests of
the dynamics of the gravitational field (postulate (1)). For more details and
references we refer the reader to [2] or [3].

\vglue 6mm

\noindent {\bf 2. Experimental tests of the coupling between matter and
gravity}

\vglue 4mm

The fact that the matter Lagrangian (2b) depends only on a symmetric tensor
$g_{\mu \nu} (x)$ and its first derivatives (i.e. the postulate of a ``metric
coupling'' between matter and gravity) is a strong assumption (often referred
to as the ``equivalence principle'') which has many observable consequences for
the behaviour of localized test systems embedded in given, external
gravitational fields. Indeed, using a  theorem of Fermi and Cartan [4] (stating
the existence of coordinate systems such that, along any given time-like curve,
the metric components can be set to their Minkowski values, and their first
derivatives made to vanish), one derives from the postulate (2) the following
observable consequences:

\item{C$_1$} : Constancy of the ``constants'' : the outcome of local
non-gravitational expe\-riments, referred to local standards, depends only on
the values of the coupling constants and mass scales entering the Standard
Model. [In particular, the cosmological evolution of the universe at large has
no influence on local experiments].

\item{C$_2$} : Local Lorentz invariance : local non-gravitational experiments
exhibit no preferred directions in spacetime [i.e. neither spacelike ones
(isotropy), nor timelike ones (boost invariance)].

\item{C$_3$} : ``Principle of geodesics'' and universality of free
fall : small, electrically neutral, non self-gravitating bodies follow
geodesics of the external spacetime $(V,g)$. In particular, two test bodies
dropped at the same location and with the same velocity in an external
gravitational field fall in the same way, independently of their masses and
compositions.

\item{C$_4$} : Universality of gravitational redshift : when
intercompared by means of electromagnetic signals, two identically constructed
clocks located at two different positions in a static external Newtonian
potential $U (\hbox{\bf x})$ exhibit, independently of their nature and
constitution, the difference in clock rate:
$$
{\tau_1 \over \tau_2} = {\nu_2 \over \nu_1} = 1 + {1 \over c^2} \ [U(\hbox{\bf
x}_1) - U(\hbox{\bf x}_2)] + O \left( {1 \over c^4}\right) . \eqno (4)
$$

Many experiments or observations have tested the observable consequen\-ces $C_1
- C_4$ and found them to hold within the experimental errors. Many sorts of
data
(from spectral lines in distant galaxies to a natural fission reactor
phenomenon which took place in Gabon two billion years ago) have been used to
set limits on a possible time variation of the basic coupling constants of the
Standard Model. The best results concern the fine-structure constant $\alpha$
for the variation of which a conservative upper bound is [5]
$$
\left\vert {\dot{\alpha} \over \alpha} \right\vert < 10^{-15} \ {\rm yr}^{-1} ,
\eqno (5) $$
which is much smaller than the cosmological time scale $\sim 10^{-10} \
{\rm yr}^{-1}$.

Any ``isotropy of space'' having a direct effect on the energy levels of atomic
nuclei has been constrained to the impressive $10^{-27}$ level [6]. The
universality of free fall has been verified at the $3\times 10^{-12}$ level for
laboratory bodies [7] and at the $10^{-12}$ level for the gravitational
accelerations of the Moon and the Earth toward the Sun [8]. The ``gravitational
redshift'' of clock rates given by eq. (4) has been verified at the $10^{-4}$
level by comparing a hydrogen-maser clock flying on a rocket up to an altitude
$\sim 10 \ \! 000$ km to a similar clock on the ground.

In conclusion, the main observable consequences of the Einsteinian postulate
(2) concerning the coupling between matter and gravity (``equivalence
principle'') have been verified with high precision by all experiments to date.
The traditional view (first put forward by Fierz [10]) is that the extremely
high precision of free fall experiments ($10^{-12}$ level) strongly suggests
that the coupling between matter and gravity is exactly of the ``metric'' form
(2), but leaves open possibilities more general than eq. (1) for the
spin-content and dynamics of the fields mediating the gravitational
interaction. We shall provisionally adopt this conclusion to discuss the tests
of the other Einsteinian postulate, eq. (1). However, we shall emphasize at the
end that recent theoretical findings suggest a rather different view.

\vglue 6mm

\noindent {\bf 3. Tests of the dynamics of the gravitational field in the weak
field regime}

\vglue 4mm

Let us now consider the experimental tests of the dynamics of the gravitational
field, defined in General Relativity by the action functional (1). Following
first the traditional view, it is convenient to enlarge our framework by
embedding General Relativity within the class of the most natural relativistic
theories of gravitation which satisfy exactly the matter-coupling tests
discussed above while differing in the description of the degrees of freedom of
the gravitational field. This class of theories are the metrically-coupled
tensor-scalar theories, first introduced by Fierz [10] in a work where he
noticed that the class of non-metrically-coupled tensor-scalar theories
previously introduced by Jordan [11] would generically entail unacceptably
large violations of the consequence C$_1$. [The fact that it would, by the same
token, entail even larger violations of the consequence C$_3$ was, probably,
first noticed by Dicke in subsequent work]. The metrically-coupled (or
equivalence-principle respecting) tensor-scalar theories are defined by keeping
the postulate (2), but replacing the postulate (1) by demanding that the
``physical'' metric $g_{\mu \nu}$ be a composite object of the form
$$
g_{\mu \nu} = A^2 (\varphi) \ g_{\mu \nu}^* , \eqno (6)
$$
where the dynamics of the ``Einstein'' metric $g_{\mu \nu}^*$ is defined by the
action functional (1) (written with the replacement $g_{\mu \nu} \rightarrow
g_{\mu \nu}^*$) and where $\varphi$ is a massless scalar field. [More
generally, one can consider several massless scalar fields, with an action
functional of the form of a general nonlinear $\sigma$ model [12]]. In other
words, the action functional describing the dynamics of the spin 2 and spin 0
degrees of freedom contained in this generalized theory of gravitation reads
$$
S_{\rm gravitational} \ [g_{\mu \nu}^* ,\varphi ] = {c^4 \over 16\pi \ G_*}
\int
{d^4 x \over c} \ \sqrt{g_*} \ \left[R(g_*) - 2g_*^{\mu \nu} \ \partial_{\mu} \
\varphi \ \partial_{\nu} \ \varphi \right] . \eqno (7)
$$
Here, $G_*$ denotes some bare gravitational coupling constant. This class of
theories contains an arbitrary function, the ``coupling function''
$A(\varphi)$.
When $A(\varphi) = {\rm const.}$, the scalar field is not coupled to matter and
one falls back (with suitable boundary conditions) on Einstein's theory. The
simple, one-parameter subclass $A(\varphi) = \exp (\alpha_0 \ \varphi)$ with
$\alpha_0 \in \Rb$ is the Jordan-Fierz-Brans-Dicke theory. In the general
case, one can define the (field-dependent) coupling strength of $\varphi$ to
matter by $$\alpha (\varphi) \equiv {\partial \ln A(\varphi) \over \partial
\varphi} . \eqno (8)
$$
It is possible to work out in detail the observable consequences of
tensor-scalar theories and to contrast them with the general relativistic case
(see ref. [12] for a recent treatment).

Let us now consider the experimental tests of the dynamics of the
gravitational field that can be performed in the solar system. Because the
planets move with slow velocities $(v/c \sim 10^{-4})$ in a very weak
gravitational potential $(U/c^2 \sim (v/c)^2 \sim 10^{-8})$, solar system tests
allow us only to probe the quasi-static, weak-field regime of relativistic
gravity (technically called the ``post-Newtonian'' limit). In this limit all
solar-system gravitational experiments, interpreted within tensor-scalar
theories, differ from Einstein's predictions only through the appearance of two
``post-Einstein'' parameters $\overline{\gamma}$ and $\overline{\beta}$
(related to the usually considered post-Newtonian parameters through
$\overline{\gamma} \equiv \gamma -1$, $\overline{\beta} \equiv \beta -1$). The
parameters $\overline{\gamma}$ and $\overline{\beta}$ vanish in General
Relativity, and are given in tensor-scalar theories by
$$\overline{\gamma} = -2 \ {\alpha_0^2 \over 1+\alpha_0^2} , \eqno (9{\rm a})
$$
$$\overline{\beta} = +{1 \over 2} \ {\beta_0 \ \alpha_0^2 \over
(1+\alpha_0^2)^2} , \eqno (9{\rm b})$$
where $\alpha_0 \equiv \alpha (\varphi_0)$, $\beta_0 \equiv \partial \alpha
(\varphi_0) / \partial \varphi_0$; $\varphi_0$ denoting the
cosmologically-determined value of the scalar field far away from the solar
system. Essentially, the parameter $\overline{\gamma}$ depends only
on the linearized structure of the gravitational theory (and is a direct
measure of its field content, i.e. whether it is pure spin 2 or contains an
admixture of spin 0), while the parameter $\overline{\beta}$
parametrizes some of the quadratic nonlinearities in the field equations (cubic
vertex of the gravitational field). All currently performed gravitational
experiments in the solar system, including perihelion advances of pla\-netary
orbits, the bending and delay of electromagnetic signals passing near the Sun,
and very accurate range data to the Moon obtained by laser echoes, are
compatible with the general relativistic predictions $\overline{\gamma} = 0
=\overline{\beta}$ and give upper bounds on both $\left\vert
\overline{\gamma} \right\vert$ and $\left\vert \overline{\beta} \right\vert$
(i.e. on possible fractional deviations from General Relativity) of order
$10^{-3}$ [8], [13]. Recently, the parametrization of the weak-field deviations
between generic tensor-multi-scalar theories and Einstein's theory has been
extended to the post-post-Newto\-nian order [14]. Only two post-post-Einstein
parameters, representing a deeper layer of structure of the gravitational
interaction, show up. See [14] for a detailed discussion, including the
consequences for the interpretation of future, higher-precision solar-system
tests.

\vglue 6mm

\noindent {\bf 4. Tests of the dynamics of the gravitational field in the
strong field regime}

\vglue 4mm

In spite of the diversity, number and often high precision of solar system
tests, they have an important qualitative weakness : they probe neither the
radiation pro\-perties nor the strong-field aspects of relativistic gravity.
Fortunately, the discovery [15] and continuous observational study of pulsars
in gravitationally bound binary orbits has opened up an entirely new testing
ground for relativistic gravity, giving us an experimental handle on the
regime of strong and/or radiative gravitational fields.

The fact that binary pulsar data allow one to probe the propagation properties
of the gravitational field is well known. This comes directly from the fact
that the finite velocity of propagation of the gravitational interaction
between the pulsar and its companion generates damping-like terms in the
equations of motion, i.e. terms which are directed against the velocities.
[This can be understood heuristically by considering that the finite velocity
of propagation must cause the gravitational force on the pulsar to make an
angle with the instantaneous position of the companion [16], and was verified
by a careful derivation of the general relativistic equations of motion of
binary systems of compact objects [17]]. These damping forces cause the binary
orbit to shrink and its orbital period $P_b$ to decrease. The remarkable
stability of the pulsar clock, together with the cleanliness of the binary
pulsar system, has allowed Taylor and collaborators to measure the secular
orbital period decay $\dot{P}_b \equiv dP_b / dt$ [18], thereby giving us a
direct experimental probe of the damping terms present in the equations of
motion. Note that, contrary to what is commonly stated, the link between the
observed quantity $\dot{P}_b$ and the propagation properties of the
gravitational interaction is quite direct. [It appears indirect only when one
goes through the common but unnecessary detour of a heuristic reasoning based
on the consideration of the energy lost in the gravitational waves emitted at
infinity].

The fact that binary pulsar data allow one to probe strong-field aspects of
re\-lativistic gravity is less well known. The a priori reason for saying that
they should is that the surface gravitational potential of a neutron star $Gm
/ c^2 R \simeq 0.2$ is a mere factor 2.5 below the black hole limit (and a
factor $\sim 10^8$ above the surface potential of the Earth). It has been
recently shown [19] that a self-gravity as strong as that of a neutron star
can naturally (i.e. without fine tuning of parameters) induce order-unity
deviations from general relativistic predictions in the orbital dynamics of a
binary pulsar thanks to the existence of nonperturbative strong-field effects
in tensor-scalar theories. [The adjective ``nonperturbative'' refers here to
the fact that this phenomenon is nonanalytic in the coupling strength of
scalar field, eq. (8), which can be as small as wished in the weak-field
limit]. As far as we know, this is the first example where large deviations
from General Relativity, induced by strong self-gravity effects, occur in a
theory which contains only positive energy excitations and whose
post-Newtonian limit can be arbitrarily close to that of General Relativity.

A comprehensive account of the use of binary pulsars as laboratories for
testing strong-field gravity has been recently given [20]. Two complementary
approaches can be pursued : a phenomenological one (``Parametrized
Post-Keplerian'' formalism), or a theory-dependent one [12], [20].

The phenomenological analysis of binary pulsar timing data consists in fitting
the observed sequence of pulse arrival times to the generic DD timing formula
[21] whose functional form has been shown to be common to the whole class of
tensor-multi-scalar theories. The least-squares fit between the timing data
and the parameter-dependent DD timing formula allows one to measure, besides
some ``Keplerian'' parameters (``orbital period'' $P_b$, ``eccentricity''
$e$,$\ldots$), a maximum of eight ``post-Keplerian'' para\-meters : $k,\gamma
,\dot{P}_b ,r,s,\delta_{\theta} ,\dot e$ and $\dot x$. Here, $k\equiv
\dot{\omega} P_b / 2\pi$ is the fractional periastron advance per orbit,
$\gamma$ a time dilation parameter (not to be confused with its post-Newtonian
namesake), $\dot{P}_b$ the orbital period derivative mentioned above,
and $r$ and $s$ the ``range'' and ``shape'' parameters of the gravitational
time delay caused by the companion. The important point is that the
post-Keplerian parameters can be measured without assuming any specific theory
of gravity. Now, each specific relativistic theory of gravity predicts that,
for instance, $k,\gamma, \dot{P}_b ,r$ and $s$ (to quote parameters that have
been successfully measured from some binary pulsar data) are some
theory-dependent functions of the (unknown) masses $m_1 ,m_2$ of the pulsar
and its companion. Therefore, in our example, the five simultaneous
phenomenological measurements of $k,\gamma ,\dot{P}_b ,r$ and $s$ determine,
for each given theory, five corresponding theory-dependent curves in the  $m_1
-m_2$ plane (through the 5 equations $k^{\rm measured} = k^{\rm theory} (m_1
,m_2 )$, etc$\ldots$). This yields three $(3=5-2)$ tests of the specified
theory, according to whether the five curves meet at one point in the mass
plane, as they should. In the most ge\-neral (and optimistic) case, discussed
in
[20], one can phenomenologically analyze both timing data and pulse-structure
data (pulse shape and polarization) to extract up to nineteen post-Keplerian
parameters. Simultaneous measurement of these 19 para\-meters in one binary
pulsar system would yield 15 tests of relativistic gravity (where one must
subtract 4 because, besides the two unknown masses $m_1 ,m_2$, generic
post-Keplerian parameters can depend upon the two unknown Euler angles
determining the direction of the spin of the pulsar). The theoretical
significance of these tests depends upon the physics lying behind the
post-Keplerian parameters involved in the tests. For instance, as we said
above, a test involving $\dot{P}_b$ probes the propagation (and helicity)
properties of the gravitational interaction. But a test involving, say,
$k,\gamma ,r$ or $s$ probes (as shown by combining the results of [12] and
[19]) strong self-gravity effects independently of radiative effects.

Besides the phenomenological analysis of binary pulsar data, one can also
adopt a theory-dependent methodology [12], [20]. The idea here is to work from
the start within a certain finite-dimensional ``space of theories'', i.e.
within a specific class of gravitational theories labelled by some theory
parameters. Then by fitting the raw pulsar data to the predictions of the
considered class of theories, one can determine which regions of theory-space
are compatible (at say the 90\% confidence level) with the available
experimental data. This method can be viewed as a strong-field genera\-lization
of the parametrized post-Newtonian formalism [2] used to analyze solar-system
experiments. In fact, under the assumption that strong-gravity effects in
neutron stars can be expanded in powers of the ``compactness'' $c_A \equiv -2 \
\partial \ {\rm ln} \ m_A / \partial \ {\rm ln} \ G \sim G \ m_A / c^2 \ R_A$,
Ref. [12] has shown that the observable predictions of generic
tensor-multi-scalar theories could be parametrized by a sequence of ``theory
parameters'', $$
\overline{\gamma} \ , \ \overline{\beta} \ , \ \beta_2 \ , \ \beta' \ ,
\ \beta'' \ , \ \beta_3 \ , \ (\beta \beta') \ldots \eqno (10)
$$
representing deeper and deeper layers of structure of the relativistic
gravitational interaction beyond the first-order post-Newtonian level
parametrized by $\overline{\gamma}$ and $\overline{\beta}$ (the second layer
$\beta_2 ,\beta'$ parametrizing the second-order post-Newtonian level [14],
etc$\ldots$). A specific two-parameter subclass of tensor-bi-scalar theories
$T(\beta' ,\beta'')$ has been given special consideration [12], [20].

After having reviewed the theory of pulsar tests, let us briefly summarize the
current experimental situation. Concerning the first discovered binary pulsar
PSR$1913+16$ [15], it has been possible to measure with accuracy the three
post-Keplerian para\-meters $k, \gamma$ and $\dot{P}_b$. From what was said
above, these three simultaneous measurements yield {\it one} test of
gravitation theories. After subtracting a small ($\sim 10^{-14}$ level in
$\dot{P}_b$ !), but significant, perturbing effect caused by the Galaxy [22],
one finds that General Relativity passes this $( k-\gamma
-\dot{P}_b )_{1913+16}$ test with complete success at the $3.5\times 10^{-3}$
level [23], [18]. This beautiful confirmation of General Relativity is an
embarrassment of riches in that it probes, at the same time, the propagation
{\it and} strong-field properties of relativistic gravity ! If the timing
accuracy of PSR$1913+16$ could improve by a significant factor two more
post-Keplerian parameters ($r$ and $s$) would become measurable and would allow
one to probe separately the propagation and strong-field aspects [23].
Fortunately, the recent discovery of the binary pulsar PSR$1534+12$ [24] (which
is significantly stronger than PSR$1913+16$ and has a more favourably oriented
orbit) has opened a new testing ground, in which it has been possible, already
after one year of data taking, to probe strong-field gravity independently of
radiative effects. A phenomenological analysis of the timing data of
PSR$1534+12$ has allowed one to measure the four post-Keplerian parameters
$k,\gamma ,r$ and $s$ [23]. From what was said above, these four simultaneous
measurements yield {\it two} tests of strong-field gravity, without mixing of
radiative effects. General Relativity is found to pass these tests with
complete success within the measurement accuracy [23], [18]. More recently, it
has been possible to extract also the ``radiative'' parameter $\dot{P}_b$ from
the timing data of PSR$1534+12$. Again, General Relativity is found to be fully
consistent (at the current $\sim 20\%$ level) with the additional test provided
by the $\dot{P}_b$ measurement [25]. Note that this gives our second direct
experimental confirmation that the gravitational interaction propagates as
predicted by Einstein's theory. Moreover, an analysis of the pulse shape of
PSR$1534+12$ has shown that the misalignment between the spin vector of the
pulsar and the orbital angular momentum was greater than $8\up{0}$ [20]. This
opens the possibility that this system will soon allow one to test the spin
precession induced by gravitational spin-orbit coupling.

To end this brief summary, let us mention that a comprehensive
theory-dependent analysis of all available pulsar data has been performed, and
has led to significant bounds on the strong-field parameters $\beta' ,\beta''$
[23]. In spite of the impressive agreement between the predictions of General
Relativity in the strong-field regime and all current binary pulsar data, the
number and precision of present strong-field tests is still rather small, and
it is important to continue obtaining and/or improving such tests, especially
in view of the results of [19] which prove that such tests are logically
independent from solar-system tests.

For a general review of the use of pulsars as physics laboratories the
reader can consult Ref. [26].

\vglue 6mm

\noindent {\bf 5. Was Einstein 100\% right ?}

\vglue 4mm

Summarizing the experimental evidence discussed above, we can say that
Einstein's postulate of a pure metric coupling between matter and gravity
(``equivalence principle'') appears to be, at least, $99.999 \ \! 999 \ \! 999
\
\! 9\%$ right (because of universality-of-free-fall experiments), while
Einstein's postulate (1) for the field content and dynamics of the
gravitational field appears to be, at least, $99.9\%$ correct both in the
quasi-static-weak-field limit appropriate to solar-system experiments, and in
the radiative-strong-field regime explored by binary pulsar experiments. Should
one apply Ockham's razor and decide that Einstein must have been $100\%$ right,
and then stop testing General Relativity ? My answer is definitely, no !

First, one should continue testing a basic physical theory such as Ge\-neral
Relativity to the utmost precision available simply because it is one of the
essential pillars of the framework of physics. Second, some very crucial
qualitative features of General Relativity have not yet been verified : in
particular the existence of black holes, and the direct detection on Earth of
gravitational waves. [Hopefully, the LIGO/VIRGO network of interferometric
detectors will observe gravitational waves early in the next century].

Last, some recent theoretical findings suggest that the current level of
precision of the experimental tests of gravity might be naturally (i.e.
without fine tu\-ning of parameters) compatible with Einstein being actually
only 50\%, or even 33\% \break right ! By this we mean that the correct theory
of gravity could involve, on the same fundamental level as the Einsteinian
tensor field $g_{\mu \nu}^*$, a massless scalar field $\varphi$ which could
(``50\% right'') or could not (``33\% right'') be coupled to matter in keeping
with the equivalence principle (2).

Let us first follow the traditional view (initiated by Fierz and enshrined by
Dicke, Nordtvedt and Will [2]) that the $10^{-12}$ level of testing of the
universality of free fall is so impressive that one should apply Ockham's
razor for what concerns the equivalence principle (2), but not yet for the
first postulate (1) which is tested only at the $10^{-3}$ level. If then we
impose the usual consistency requirements of field theory (absence of
algebraic inconsistencies, discontinuities in the degree-of-freedom content,
causa\-lity problems, negative-energy excitations,$\ldots$) we are uniquely led
to considering only the class of metrically-coupled tensor-multi-scalar
theories.

It has been shown that the (positive-energy) multi-scalar case did not bring
essentially new features with respect to the mono-scalar case [12]. We
therefore limit our discussion to the simplest tensor-scalar theories defined
in section 3 above.

Because of the authority of Dicke, it has become common, when discussing
equivalence-principle respecting tensor-scalar theories, to restrict one's
attention to the one-parame\-ter subclass characterized by the coupling
function
$A(\varphi) =\exp (\alpha_0 \ \varphi)$. [This defines the
Jordan-Fierz-Brans-Dicke theory, introduced by Fierz [10] as the one-parameter,
metrically-coupled subclass of the two-parameter theory of Jordan [11]]. For
many years, the low precision of solar-system relativistic tests (and the
possibility of the Sun having a sizable quadrupole moment) did not put strong
constraints on the coupling constant $\alpha_0^2 = (2\omega +3)^{-1}$, leaving
open the possibility that $\alpha_0$ be of order unity (as expected if
$\varphi$ is to be a fundamental field, on the same footing as $g_{\mu
\nu}^*$). The result of the Viking relativistic time delay experiment [13]
(namely $\alpha_0^2 < 10^{-3}$) shattered this idea, and cast a serious doubt
on the a priori plausibility of tensor-scalar theories. In my view, the
situation has been significantly transformed by a recent work [27] which found
that the general class of (metrically-coupled) tensor-scalar theories, with
arbitrary coupling function $A(\varphi)$, generically contain an ``attractor
mechanism'' toward General Relativity. More precisely, as soon as the function
$a(\varphi) \equiv \ {\rm ln} \ A(\varphi)$ admits a minimum, the cosmological
evolution tends to drive the cosmic value (or Vacuum Expectation Value, VEV)
$\varphi_0$ of the scalar field toward a value where $a(\varphi)$ reaches a
minimum, i.e. a value where the effective coupling strength of the scalar field
$\alpha (\varphi) =\partial \ a(\varphi) / \partial \varphi$, eq. (8),
vanishes. Seen from this point of view, it is na\-tural to expect (in a wide
class of tensor-scalar theories) that the present value of the scalar coupling
strength $\alpha_0 = \alpha (\varphi_0 )$ be much smaller than unity. [Note
that the Jordan-Fierz-Brans-Dicke theory, with $a(\varphi )= \alpha_0 \
\varphi$, does not belong to the wide class of ``GR attracting'' theories].
Analytical estimates of the efficiency of the cosmological attractor mechanism
suggest that a natural level for the expected present deviations from General
Relativity is
$$
\alpha_0^2 \ \build{>}_{\sim}^{} \ \Omega^{-3/2} \times 10^{-7} , \eqno (11)
$$
where $\Omega = \rho^{\rm matter} / \rho^{\rm critical}$ is the usual
dimensionless measure of the average mass density in the universe. The
estimate (11) shows that the present agreement at the $10^{-3}$ level between
General Relativity and experiment might be a red herring. This gives a new
motivation for experiments which push beyond the precision of relativity
tests, such as Stanford's gyroscope experiment (Gravity Probe B) which aims at
the level $\overline{\gamma} \sim \alpha_0^2 \sim 10^{-5}$.

Let us however draw back and question the traditional view which led us to
restrict our attention to equivalence-principle respecting theories. This view
stemmed from the work of Fierz who noticed that the most general tensor-scalar
theory of Jordan would strongly violate the equivalence principle. Fierz's
proposal to modify the scalar couplings so as to be in keeping with the
postulate (2) was an ad hoc way of preventing too violent a contradiction with
experiment. However, if we ask in turn why Jordan had been led to considering
theories containing equivalence-principle violating couplings $\propto \
e^{\alpha \varphi} \ F_{\mu \nu} \ F^{\mu \nu}$, between $\varphi$ and gauge
fields, the answer is that such couplings were necessary consequences of the
Kaluza-Klein unification programme that Jordan was developing. And if we ask
what kind of couplings are predicted by all the modern versions of the
programme of unifying gravity with the other interactions (generalized
Kaluza-Klein, extended supergravity, string theory) the answer is that they
generically predict the existence of massless scalar fields coupled in an
equivalence-principle-violating way. At this juncture, one would be tempted to
conclude that this suggests that all the eventual scalar partners of the
Einsteinian tensor field must acquire a mass and thereby bring only
negligible, exponentially small corrections $\propto \ \exp (-mr/\hbar c)$ to
the general relativistic predictions concerning low-energy gravitational
effects. An alternative possibility, which gives a new motivation for tes\-ting
the equivalence principle, has been recently proposed [28] : string-loop
effects (i.e. quantum corrections induced by worldsheets of arbitrary genus in
intermediate string states) may modify the low-energy, Kaluza-Klein-type,
matter couplings of the massless scalars present in string theory (dilaton or
moduli fields) in such a manner that, through a genera\-lization of the
attractor mechanism discussed above, the va\-cuum expectation values of the
scalar fields be cosmologically driven towards values where they decouple
from matter. For such a ``least coupling principle'' to hold, the coupling
functions of the scalar field(s) must exhibit certain properties of
universality. More precisely, the most general low-energy couplings induced by
string-loop effects will be such that the various terms on the right-hand side
of eq. (2b) will be multiplied by several different functions of the scalar
field(s) : say a factor $B_F (\varphi)$ in factor of the kinetic terms of the
gauge fields, a factor $B_{\psi} (\varphi)$ in factor of the Dirac kinetic
terms, etc$\ldots$ It has been shown in [28] that if the various coupling
functions $B_i (\varphi)$, $i=F,\psi ,\ldots$, all admit an extremum (which
must be a maximum for the ``leading'' $B_i$) at some common value $\varphi_m$
of $\varphi$, the cosmological evolution of the coupled tensor-scalar-matter
system will drive $\varphi$ towards the value $\varphi_m$, at which $\varphi$
decouples from matter. As suggested in [28] a natural way in which the
required conditions could be satisfied is through the existence of a discrete
symmetry in scalar space. [For instance, a symmetry under $\varphi \rightarrow
-\varphi$ would guarantee that all the scalar coupling functions reach an
extremum at the self-dual point $\varphi_m =0$]. The existence of such
symmetries have been proven for some of the scalar fields appearing in string
theory (target-space duality for the moduli fields) and conjectured for others
($S$-duality for the dilaton). This gives us some hope that the mechanism of
[28] could apply and thereby naturally reconcile the existence of massless
scalar fields with experiment. Indeed, a study of the efficiency of attraction
of $\varphi$ towards $\varphi_m$ [which happens to be generically larger than
in the simple case of Ref. [27], which led to eq. (11), because of the steep
dependence of all the physical mass scales upon the gauge coupling function
$B_F (\varphi)$] estimates that the present vacuum expectation value
$\varphi_0$ of the scalar field would differ (in a rms sense) from $\varphi_m$
by
$$
\varphi_0 - \varphi_m \sim 2.75 \times 10^{-9} \times \kappa^{-3} \
\Omega^{-3/4} \ \Delta \varphi \eqno (12)
$$
where $\kappa$ denotes the curvature of ${\rm ln} \ B_F (\varphi)$ around the
maximum $\varphi_m$ and $\Delta \varphi$ the deviation $\varphi - \varphi_m$
at the beginning of the (classical) radiation era. Equation (12) predicts the
existence, at the present cosmological epoch, of many small, but non zero,
deviations from General Relativity proportional to the {\it square} of
$\varphi_0 -\varphi_m$. This provides a new incentive for trying to improve by
several orders of magnitude the various experimental tests of Einstein's
equivalence principle, i.e. of the consequences $C_1 - C_4$ recalled above.
For instance, it would be interesting to improve the direct experimental
bounds on the secular change of the fine-structure constant $\alpha$ by
comparing clocks based on atomic transitions having different dependences on
$\alpha$. It seems, however, that the most sensitive way to look for a small
residual violation of the equivalence principle is to perform improved tests
of the universality of free fall. The mechanism of Ref. [28] suggests a
specific composition-dependence of the residual differential acceleration of
free fall and estimates that a non-zero signal could exist at the very small
level
$$
\left( {\Delta a \over a} \right)_{\rm rms}^{\rm max} \sim 1.36 \times
10^{-18} \ \kappa^{-4} \ \Omega^{-3/2} \ (\Delta \varphi)^2 , \eqno (13)
$$
where $\kappa$ is expected to be of order unity (or smaller, leading to a
larger signal, in the case where $\varphi$ is a modulus rather than the
dilaton). Let us emphasize again that the strength of the cosmological scenario
considered here as counterargument to applying Ockham's razor lies in the fact
that the very small number on the right-hand side of eq. (13) has been derived
without any fine tuning or use of small parameters. The estimate (13) gives
added significance to the project of a Satellite Test of the Equivalence
Principle (nicknamed STEP, and currently studied by ESA, NASA and CNES) which
aims at probing the universality of free fall of pairs of test masses orbiting
the Earth at the $10^{-17}$ level [29].

\vglue 6mm

\noindent {\bf References}

\vglue 4mm

\item{[1]} R.P. Feynman, {\it Lectures on Gravitation}, unpublished lecture
notes (1962-1963) prepared by
F.B. Morinigo and W.G. Wagner (California Institute of Technology, 1971);

S. Weinberg, Phys. Rev. {\bf 138} (1965) B988,

V.I. Ogievetsky and I.V. Polubarinov, Ann. Phys. N.Y. {\bf 35} (1965) 167;

W. Wyss, Helv. Phys. Acta {\bf 38} (1965) 469;

S. Deser, Gen. Rel. Grav. {\bf 1} (1970) 9;

D.G. Boulware and S. Deser, Ann. Phys. N.Y. {\bf 89} (1975) 193;

J. Fang and C. Fronsdal, J. Math. Phys. {\bf 20} (1979) 2264;

R.M. Wald, Phys. Rev. D {\bf 33} (1986) 3613;

C. Cutler and R.M. Wald, Class. Quantum Grav. {\bf 4} (1987) 1267;

R.M. Wald, Class. Quantum Grav. {\bf 4} (1987) 1279.

\item{[2]} C.M. Will, {\it Theory and Experiment in Gravitational Physics},
2nd edition (Cambridge University Press, Cambridge, 1992); and Int. J. Mod.
Phys. D {\bf 1} (1992) 13.

\item{[3]} T. Damour, {\it Gravitation and Experiment} in {\it Gravitation and
Quantizations}, eds B. Julia and J. Zinn-Justin, Les Houches, Session LVII
(Elsevier, Amsterdam, 1994).

\item{[4]} E. Fermi, Atti Accad. Naz. Lincei Cl. Sci. Fis. Mat. \& Nat. {\bf
31} (1922) 184 and 306;

\item{ \ } E. Cartan, {\it Le\c cons sur la G\'eom\'etrie des Espaces de
Riemann} (Gauthier-Villars, Paris, 1963).

\item{[5]} P. Sisterna and H. Vucetich, Phys. Rev. D {\bf 41} (1990) 1034.

\item{[6]} J.D. Prestage et al., Phys. Rev. Lett. {\bf 54} (1985) 2387;

S.K. Lamoreaux et al., Phys. Rev. Lett. {\bf 57} (1986) 3125;

T.E. Chupp et al., Phys. Rev. Lett. {\bf 63} (1989) 1541.

\item{[7]} Y. Su et al., Phys. Rev. D {\bf 50} (1994) 3614.

\item{[8]} J.O. Dickey et al., Science {\bf 265} (1994) 482; see also J.G.
Williams, X.X. Newhall and J.O. Dickey, these proceedings.

\item{[9]} R.F.C. Vessot and M.W. Levine, Gen. Rel. Grav. {\bf 10} (1978)
181;

R.F.C. Vessot et al., Phys. Rev. Lett. {\bf 45} (1980) 2081.

\item{[10]} M. Fierz, Helv. Phys. Acta {\bf 29} (1956) 128.

\item{[11]} P. Jordan, Nature {\bf 164} (1949) 637; {\it Schwerkraft und
Weltall} (Vieweg, Braunschweig, 1955).

\item{[12]} T. Damour and G. Esposito-Far\`ese, Class. Quant. Grav. {\bf 9}
(1992) 2093.

\item{[13]} R.D. Reasenberg et al., Astrophys. J. {\bf 234} (1979) L219.

\item{[14]} T. Damour and G. Esposito-Far\`ese, to be published.

\item{[15]} R.A. Hulse and J.H. Taylor, Astrophys. J. Lett. {\bf 195} (1975)
L51; see also the 1993 Nobel lectures in physics of Hulse (pp. 699-710) and
Taylor (pp. 711-719) in Rev. Mod. Phys. {\bf 66}, n\up{0}3 (1994).

\item{[16]} P.S. Laplace, {\it Trait\'e de M\'ecanique C\'eleste}, (Courcier,
Paris, 1798-1825), Second part : book 10, chapter 7.

\item{[17]} T. Damour and N. Deruelle, Phys. Lett. A {\bf 87} (1981) 81;

T. Damour, C.R. Acad. Sci. Paris {\bf 294} (1982) 1335;

\item{ \ } T. Damour, in {\it Gravitational Radiation}, eds N. Deruelle and T.
Piran (North-Holland, Amsterdam, 1983) pp 59-144.

\item{[18]} J.H. Taylor, Class. Quant. Grav. {\bf 10} (1993) S167
(Supplement 1993) and refe\-rences therein; see also J.H. Taylor's Nobel
lecture quoted in [15].

\item{[19]} T. Damour and G. Esposito-Far\`ese, Phys. Rev. Lett. {\bf 70}
(1993) 2220.

\item{[20]} T. Damour and J.H. Taylor, Phys. Rev. D. {\bf 45} (1992) 1840.

\item{[21]} T. Damour and N. Deruelle, Ann. Inst. H. Poincar\'e {\bf 43}
(1985) 107 and {\bf 44} (1986) 263.

\item{[22]} T. Damour and J.H. Taylor, Astrophys. J. {\bf 366} (1991) 501.

\item{[23]} J.H. Taylor, A. Wolszczan, T. Damour and J.M. Weisberg, Nature
{\bf 355} (1992) 132.

\item{[24]} A. Wolszczan, Nature {\bf 350} (1991) 688.

\item{[25]} A. Wolszczan and J.H. Taylor, to be published (quoted in Taylor's
Nobel lecture [15]).

\item{[26]} R.D. Blandford et al. editors, {\it Pulsars as physics
laboratories}, Phil. Trans. R. Soc. London A {\bf 341} (1992) pp 1-192; see
notably the contributions by J.H. Taylor (pp. 117-134) and by T. Damour (pp.
135-149).

\item{[27]} T. Damour and K. Nordtvedt, Phys. Rev. Lett. {\bf 70} (1993)
2217; Phys. Rev. D {\bf 48} (1993) 3436.

\item{[28]} T. Damour and A.M. Polyakov, Nucl. Phys. B {\bf 423} (1994) 532;
Gen. Rel. Grav. {\bf 26} (1994), in press.

\item{[29]} P.W. Worden, in {\it Near Zero : New Frontiers of Physics}, eds
J.D. Fairbank et al. (Freeman, San Francisco, 1988) p. 766;

\item{ \ } J.P. Blaser et al., {\it STEP Assessment Study Report}, ESA document
SCI (94)5, May 1994;

R. Bonneville, GEOSTEP, CNES document, DP/SC, April 1994.

\par}

\bye